\documentclass{ws-rv9x6}
\usepackage{subfigure}   
\usepackage{ws-rv-thm}   
\usepackage{ws-rv-van}   

\newcommand{\esp}[1]{\, e^{\,\,\textstyle {#1}}}
\makeindex

\begin{document}

\chapter[Using World Scientific's Review Volume Document Style]{Multiple Parton Interactions, inclusive and exclusive cross sections, sum rules}
\label{ra_ch1}

\author[D. Treleani and G. Calucci]{D. Treleani and G. Calucci\footnote{Now retired}}

\address{Dipartimento di
Fisica dell'Universit\`a di Trieste and INFN, Sezione di
Trieste,\\
Strada Costiera 11, Miramare-Grignano, I-34014 Trieste,
Italy.}

\begin{abstract}
In a simplified model of Multiple Parton Interactions the
inclusive cross sections, of processes with large momentum transfer exchange, acquire the statistical meaning of factorial moments of the
distribution in multiplicity of interactions, while more
exclusive cross sections, which can provide
complementary information on the interaction dynamics, become experimentally viable. Inclusive and exclusive cross sections are linked by sum rules, which can be tested experimentally.
\end{abstract}

\body

\section{Introduction}

The relevance of Multiple Parton Interactions (MPIs) in high energy hadronic collisions was apparent since from the very beginning of the description of large $p_t$ processes by pQCD\cite{Paver:1982yp, Humpert:1983pw, Mekhfi:1983az, Mekhfi:1985dv}. To recall some of the main ideas in the matter, we will review a simplified model of MPIs, where a cutoff in the transverse momentum exchange is introduced, to separate the hard from the soft component of the interaction, and one exploits the very different properties of hard and soft interactions in the transverse coordinates space. Being localised in a smaller region, the hard component of the interaction can in fact be disconnected, in such a way that each disconnected component may be treated as a independent hard sub-interaction, which constitutes the characteristic feature of the MPIs. In the actual model, each different hard sub-interaction is treated perturbatively, without however distinguishing between quarks and gluons and between different states of spin and charge. In addition only MPIs, where each disconnected hard sub-interaction is initiated by two partons, are taken into account. 

The picture of MPIs obtained in the model satisfies various non trivial properties, which will be discussed in detail.

By reviewing the kinematics of the process, in the next section we will outline the geometrical features of the Double Parton Interaction (DPI) inclusive cross section, we will introduce the "effective cross section" and we will show that interference terms between DPIs and Single Parton Interactions (SPIs) are suppressed. In the following section we will generalise the cross section to the case of an increasingly large number of MPIs and we will write the expression of the total contribution of hard interactions to the inelastic cross section in the simplest case, where all multi-parton correlations can be neglected. The case where two-body correlations play an important role will be discussed by means of a suitable functional formalism in the successive section. The possibility of introducing explicit expressions, not only for the inclusive but also for a definite set of exclusive cross sections, will be outlined in the last section, where we will discuss also the sum rules, linking inclusive and exclusive cross sections. 

\section{Double scattering}

The DPI inclusive cross section is obtained by the unitarity diagram in Fig.1, in the limit of large c.m. energy and large momentum transfer exchange in the two partonic interactions, which generate the final states with overall momenta $P$ and $Q$\cite{Paver:1982yp}, while the remnants of the hadron carry momenta $\underline{A}$ and $\underline{B}$. The soft blobs $\phi_A$ and $\phi_B$ represent the hadron bound state and the virtual lines attached to $\phi$ are characterised by transverse momenta and  off-shell scales of the order of the hadron mass. In the hadron-hadron c.m. frame, the corresponding light cone + components are of ${\cal O}\sqrt s$ in the upper part of the diagram and the light cone - components are of ${\cal O}\sqrt s$ in the lower part of the diagram, while the initial transverse components do not grow with the c.m energy $\sqrt s$. Because of momentum conservation, the overall transverse momenta of $P$ and $Q$ are of order of the transverse momenta of initial state partons, while the transverse momenta of the final state partons generated by the two hard interactions are large.

\begin{figure}[htp]
\centering
\includegraphics[width=11cm]{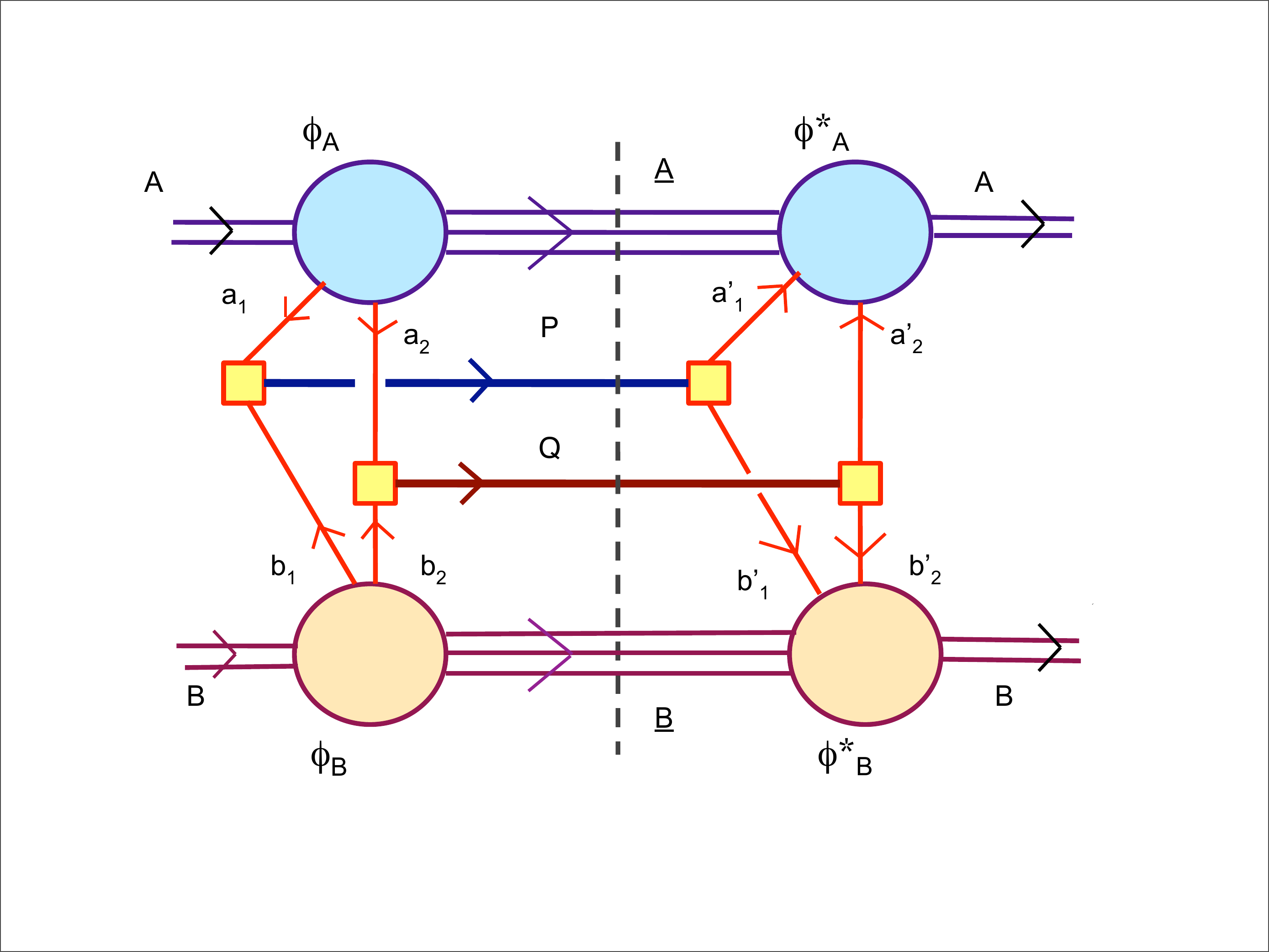}
\vspace{0cm}
\caption{Unitarity diagram for the double parton scattering cross section. $P$ and $Q$ represent the total four-momenta of the partonic states produced in the two interactions, e.g. of two pairs of large $p_t$ jets.}
\label{fig:double_scattering}
\end{figure}

\noindent The diagram is characterised by five independent loops. As loop variables one may choose

\begin{eqnarray}
P,\qquad Q,\qquad \delta=(a_1-a_2)/2,\qquad \delta'=(a_1'-a_2')/2, \qquad \underline{A}
\end{eqnarray}

\noindent Energy momentum conservation in the interaction vertices give

\begin{eqnarray}
P=a_1+b_1=a_1'+b_1',\qquad Q=a_2+b_2=a_2'+b_2'
\end{eqnarray}

\noindent and one has 

\begin{eqnarray}
&&a_1=(A-\underline{A})/2+\delta, \qquad\, b_1=P-(A-\underline{A})/2-\delta,\nonumber\\
&&a_2=(A-\underline{A})/2-\delta, \qquad\, b_2=Q-(A-\underline{A})/2+\delta,\nonumber\\
&&a_1'=(A-\underline{A})/2+\delta', \qquad b_1=P-(A-\underline{A})/2-\delta',\nonumber\\
&&a_2'=(A-\underline{A})/2-\delta', \qquad b_2=Q-(A-\underline{A})/2+\delta'.
\end{eqnarray}

\noindent One may introduce the light cone components

\begin{eqnarray}
(a_1)_-=(a_1^2+a_{1,t}^2)/(a_1)_+,\qquad  (b_1)_+=(b_1^2+b_{1,t}^2)/(b_1)_-
\end{eqnarray}

\noindent and, keeping into account that the soft blobs allow virtualities and transverse momenta only of the order of the hadronic scale $R$ (in coordinates space), one has 

\begin{eqnarray}
a_1^2\simeq a_{1,t}^2\simeq b_1^2\simeq b_{1,t}^2={\cal O}(1/R^2)
\end{eqnarray}

\noindent which implies

\begin{eqnarray}
&&(a_1)_+={\cal O}(\sqrt s), \qquad (a_1)_-={\cal O}(1/R^2\sqrt s),\nonumber\\
&&(b_1)_+={\cal O}(1/R^2\sqrt s), \qquad (b_1)_-={\cal O}(\sqrt s),
\end{eqnarray}

\begin{figure}[htp]
\centering
\includegraphics[width=11cm]{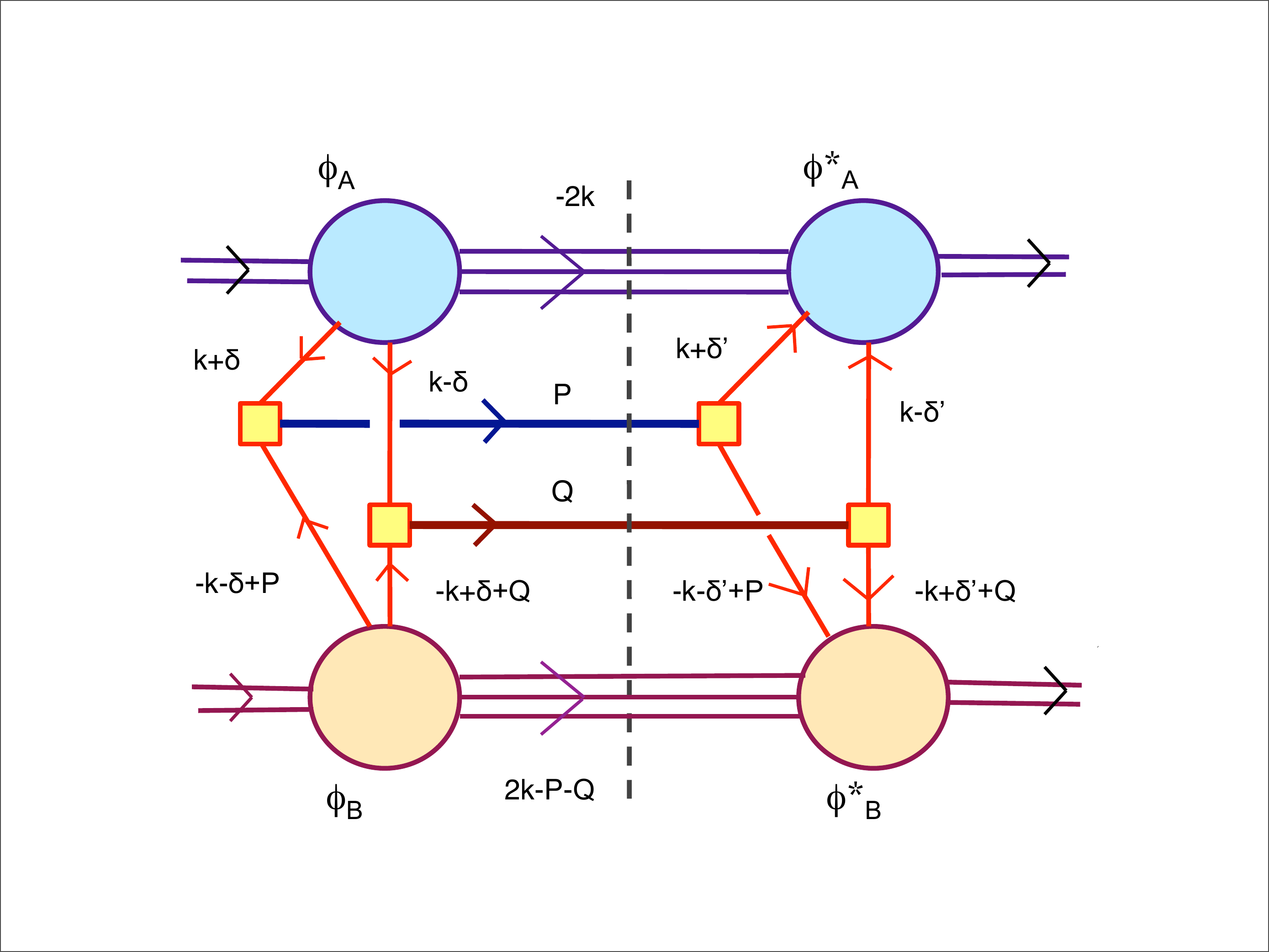}
\vspace{0cm}
\caption{Transverse momenta flow in the double parton scattering diagram}
\label{fig:transverse_double}
\end{figure}

\noindent A similar argument holds for the primed variables and for the variables with index 2. One obtains

\begin{eqnarray}
&&P_+\simeq(a_1)_+\pm{\cal O}(1/R^2\sqrt s)\simeq (a_1')_+\pm{\cal O}(1/R^2\sqrt s),\nonumber\\
&&P_-\simeq(b_1)_-\pm{\cal O}(1/R^2\sqrt s)\simeq (b_1')_-\pm{\cal O}(1/R^2\sqrt s),\nonumber\\
&&Q_+\simeq(a_2)_+\pm{\cal O}(1/R^2\sqrt s)\simeq (a_2')_+\pm{\cal O}(1/R^2\sqrt s),\nonumber\\
&&Q_-\simeq(b_2)_-\pm{\cal O}(1/R^2\sqrt s)\simeq (b_2')_-\pm{\cal O}(1/R^2\sqrt s),
\end{eqnarray}

\noindent and, as shown in Fig.\ref{fig:transverse_double}, one has

\begin{eqnarray}
&&k=(a_1+a_2)/2=(a_1'+a_2')/2,\,\,\,\delta=(a_1-a_2)/2,\,\,\,\delta'=(a_1'-a_2')/2\nonumber
\end{eqnarray}

\noindent which implies

\begin{eqnarray}
&&\delta_-\approx\delta_-'={\cal O}(1/R^2\sqrt s),\nonumber\\
&&\delta_+\approx\delta_+'=\frac{1}{2}(P-Q)_+\pm{\cal O}(1/R^2\sqrt s),
\label{delta}
\end{eqnarray}

\noindent
The loop integrations over $\delta_{\pm},\,\delta_{\pm}'$ are therefore restricted to a range of ${\cal O}(1/R^2\sqrt s)$. The integration on $\delta_-$ hence involve only the upper vertex $\phi_A$ and the propagators of the lines with momenta $a_1$ and $a_2$, whose 'minus' components are also of ${\cal O}(1/R^2\sqrt s)$.  One needs in fact to keep into account only of the kinematical variables which grow as $\sqrt s$ in the hard interaction vertices; while the lower vertex $\phi_B$ and the propagators of the lines with momenta $b_1$ and $b_2$, whose 'minus' components are of ${\cal O}(\sqrt s)$, are practically constant for variations of $\delta_-$ of ${\cal O}(1/R^2\sqrt s)$. Conversely for the integration on $\delta_+$, while similar arguments hold for the integrations on $\delta_-'$, $\delta_+'$.

The dependence on the initial state is thus through quantities like

\begin{eqnarray}
\psi(a_{1t},a_{2t},a_{1+},a_{2+},\underline A_-)\equiv\int\frac{\phi_A(a_1,a_2,\underline A)}{a_1^2a_2^2}\frac{d\delta_-}{2\pi}
\end{eqnarray}

\noindent while the values of $a_{1+},\ a_{2+},\ \underline A_-$ are determined by the final state observables $P_+,\ Q_+,\ \underline A^2$. 

All different initial state quantities $\psi$ are linked through the integrations on the transverse momenta $a_{1t},\ a_{2t}$ etc. To deal with the transverse momentum integrations one may introduce the two dimensional Fourier transform 

\begin{eqnarray}
\tilde\psi(s_{1},s_{2}&,&a_{1+},a_{2+},\underline A_-)=\cr
&&\int d^2a_{1t}d^2a_{2t}\frac{e^{i({\bf s_{1}\cdot a_{1t} +s_{2}\cdot a_{2t}})}}{(2\pi)^2}\psi(a_{1t},a_{2t},a_{1+},a_{2+},\underline A_-)
\end{eqnarray}

\noindent Keeping into account the conservation constraints in the vertices, the flow of transverse momenta may be expressed as shown in Fig.2. The discontinuity of the diagram is thus given by a Fourier integral over all independent transverse momenta. The argument of the Fourier exponential is

\begin{eqnarray}
&&\bf (k+\delta)\cdot s_{\rm 1}+(k-\delta)\cdot s_{\rm 2}+(-k-\delta+P)\cdot s_{\rm 1}'+(-k+\delta+Q)\cdot s_{\rm 2}'\nonumber\\
&&\bf-(k+\delta')\cdot s_{\rm 1}''-(k-\delta')\cdot s_{\rm 2}''-(-k-\delta'+P)\cdot s_{\rm 1}'''-(-k+\delta'+Q)\cdot s_{\rm 2}'''\nonumber\\
\\
&=&\bf (k+\delta)\cdot (S+s/{\rm 2})+(k-\delta)\cdot (S-s/{\rm 2})+(-k-\delta+P)\cdot (S'+s'/{\rm 2})\nonumber\\
&&\bf+(-k+\delta+Q)\cdot (S'-s'/{\rm 2})
-(k+\delta')\cdot (S''+s''/{\rm 2})-(k-\delta')\cdot (S''-s''/{\rm 2})\nonumber\\
&&\bf-(-k-\delta'+P)\cdot (S'''+s'''/{\rm 2})-(-k+\delta'+Q)\cdot (S'''-s'''/{\rm 2})\nonumber
\end{eqnarray}

\noindent where $\bf s_{\rm 1}, s_{\rm 2}, s_{\rm 1}', s_{\rm 2}'$ etc. are the transverse coordinates conjugate to the transverse momenta in Fig.2 and $\bf S, s, S', s'$ etc. are the centre of mass and the relative transverse coordinates of $\bf (s_{\rm 1}, s_{\rm 2})$, $\bf (s_{\rm 1}', s_{\rm 2}')$ etc. 

The integrations on transverse momenta give:

\begin{eqnarray}
&&d^2\bf\delta\to s=s'\nonumber\\
&&d^2\bf\delta'\to s''=s'''\nonumber\\
&&d^2\bf P\to S'+s'/{\rm 2}-S'''-s'''/{\rm 2}={\rm 0}\nonumber\\
&&d^2\bf Q\to S'-s'/{\rm 2}-S'''+s'''/{\rm 2}={\rm 0}\nonumber\\
&&d^2\bf k\to S=S'',
\end{eqnarray}

\noindent which implies 

\begin{eqnarray}
\bf s=s'=s''=s'''\qquad{\rm and}\qquad S=S'',\ S'=S'''
\end{eqnarray}

\noindent The cross section may hence be expressed as a function of $|\tilde\psi|^2$. More precisely one may define the two-body parton distribution 

\begin{eqnarray}
\Gamma(x_1,x_2;{\bf s})\equiv\int|\tilde\psi({\bf s}_{1},{\bf s}_{2},a_{1+},a_{2+},\underline A_-)|^2d^2{\bf S}dA_-,
\end{eqnarray}

\noindent where $x_i=(a_i)_+/A_+$, and the cross section is proportional to the convolution

\begin{eqnarray}
\int\Gamma_A(x_1,x_2;{\bf s})\Gamma_B(x_1',x_2';{\bf s})d^2{\bf s},
\end{eqnarray}

\noindent which shows that the cross section is given by the incoherent superposition of two different hard collisions, where the two interacting pairs are localised at the same relative transverse distance $\bf s$. The hard part of the interaction is therefore disconnected in two different parts, which are physically separated in the transverse coordinates space by the distance $\bf s$, which is of the order of the hadron size, namely much larger as compared with the transverse size of the two regions where the hard interactions take place. 

One should point out that this conclusion relies on the assumption that $\Gamma_A(x_1,x_2;{\bf s})$ is not singular for ${\bf s}\to0$ or, in other words, on the assumption that $\Gamma_A(x_1,x_2;{\bf s})$ is characterised by a non perturbative scale of the order of the hadron mass at small ${\bf s}$. On the other hand, a singular behaviour of $\Gamma$ at small ${\bf s}$ would necessarily induce a large relative momentum between the final states with overall momenta $P$ and $Q$, while, in the case of DPIs, the configurations one is looking for, between the states with momenta $P$ and $Q$, are those characterised by a relative momentum of the order of the hadron mass. 

The final expression of the DPI cross section $\sigma_D$ is the following\cite{Paver:1982yp} : 

\begin{eqnarray}\label{double scattering}
\frac{d\sigma_D|_{ij}}{\prod_{k=1,2}dx_kdx_k'}&=&\frac{1}{1+\delta_{ij}}\int_{p_t^c}\Gamma_A(x_1,x_2;{\bf s})\hat\sigma_i(x_1x_1')\hat\sigma_j(x_2x_2')\Gamma_B(x_1',x_2';{\bf s})d^2{\bf s}\cr
&\equiv&\frac{1}{1+\delta_{ij}}\frac{\sigma_i(x_1x_1')\sigma_j(x_2x_2')}{\sigma_{\rm eff}}
\end{eqnarray}

\noindent
where the two partonic interactions are labelled with the indices $i$, $j$ and $1+\delta_{ij}$ is the symmetry factor for the case of two identical interactions. The transverse momenta of the produced partons are integrated with the cutoff ${p_t^c}$, $\hat{\sigma}_{i,j}$ are the partonic cross sections and $\sigma_{i,j}$ the hadron-hadron inclusive cross sections. The last line here above defines the "effective cross section" $\sigma_{\rm eff}$, which summarises in a single quantity all unknowns in the process.

\section{Interference term}

\begin{figure}[htp]
\centering
\includegraphics[width=11cm]{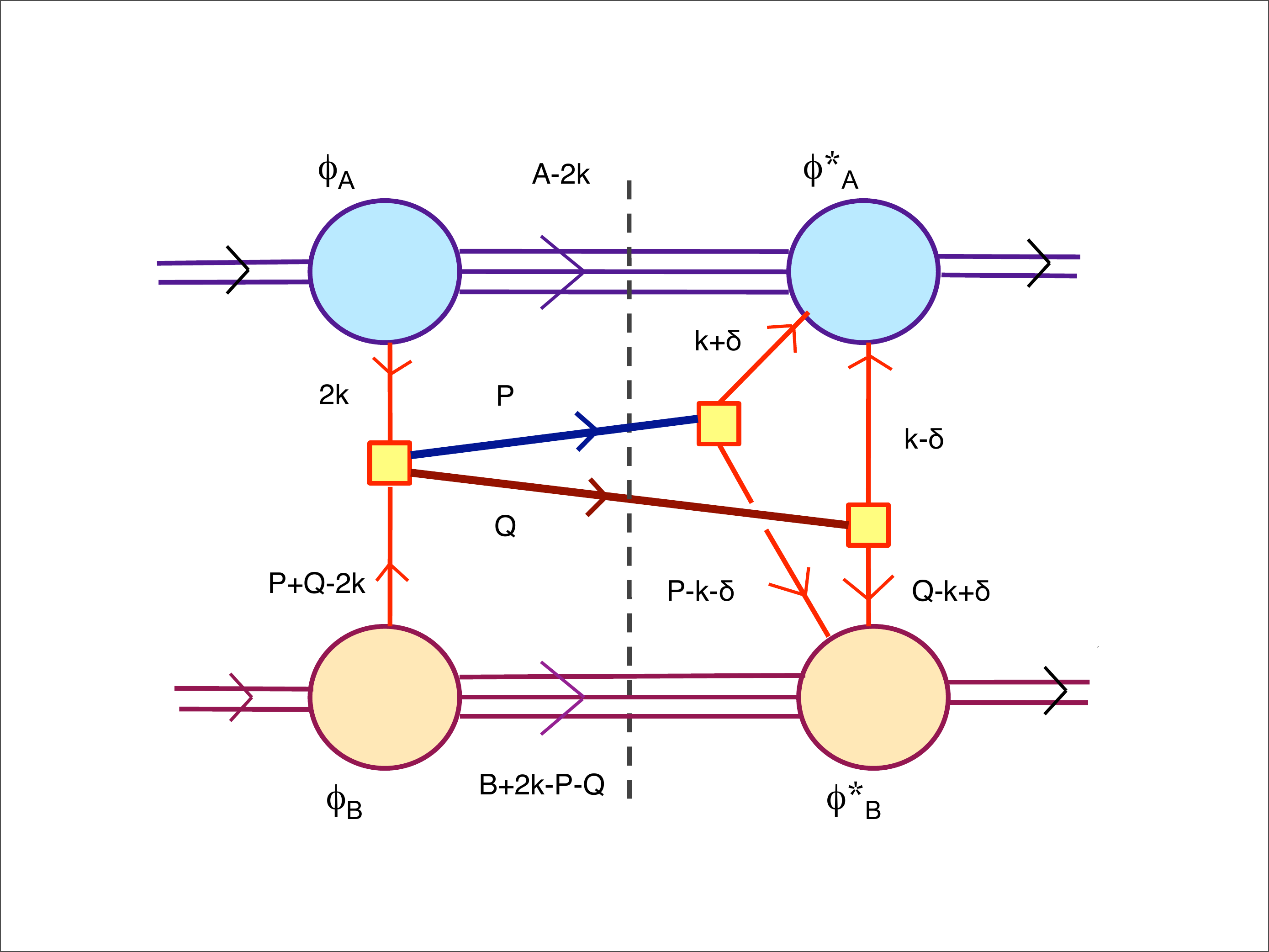}
\vspace{0cm}
\caption{Flow of momenta in a interference diagram}
\label{fig:transverse_interference}
\end{figure}

A interference diagram between a SPI and a DPI is shown in Fig.\ref{fig:transverse_interference}, with the corresponding flow of momenta. $P$ and $Q$ represent the total four-momenta of the partonic states produced in the two interactions, e.g. of two pairs of large $p_t$ jets.

It may be useful to look at the picture of the interference process in transverse space. When following the same line of reasoning of the previous section, the amplitude is expressed by the Fourier transform of the following combination, of functions of transverse coordinates of the initial state interacting partons: 

\begin{eqnarray}
\int\tilde\varphi_A({\bf s})\tilde\psi^*_A({\bf s}_1, {\bf s}_2)\tilde\varphi_B({\bf s}')\tilde\psi^*_B({\bf s}_1', {\bf s}_2')d^2{\bf s}\,d^2{\bf s}_1d^2{\bf s}_2d^2{\bf s}'d^2{\bf s}_1'd^2{\bf s}_2',
\label{transv}
\end{eqnarray}

\noindent where $\tilde\psi_{A,B}$ have been defined in the previous section, $\tilde\varphi_{A,B}$ are the analogous quantities describing the soft components on the left side of the cut in Fig.\ref{fig:transverse_interference} and the dependence on the longitudinal variables is implicit. 

The argument of the Fourier exponential is   

\begin{eqnarray}
&&2{\bf k\cdot s}+({\bf P}+{\bf Q}-2\bf k)\cdot s'-(k+\delta)\cdot s_{\rm 1}-(k-\delta)\cdot s_{\rm 2}-(P-k-\delta)\cdot s_{\rm 1}'\nonumber\\
&&\bf-(Q-k+\delta)\cdot s_{\rm 2}'
\label{transv}
\end{eqnarray}

\noindent where $\bf s,\,s_{\rm 1}, s_{\rm 2}, s', s_{\rm 1}', s_{\rm 2}'$ are the transverse coordinates conjugate to the transverse momenta shown in Fig.\ref{fig:transverse_interference}. 

\noindent The integrations on transverse momenta give

\begin{eqnarray}
d^2\bf P&\to&{\bf s'-s_{\rm 1}'=}0\nonumber\\
d^2\bf Q&\to&{\bf s'-s_{\rm 2}'=}0\nonumber\\
d^2\bf\delta&\to&{\bf s_{\rm 1}-s_{\rm 2}-s_{\rm 1}'+s_{\rm 2}'=}0\nonumber\\
d^2\bf k&\to&{\bf {\rm 2}s-{\rm 2}s'-s_{\rm 1}-s_{\rm 2}+s_{\rm 1}'+s_{\rm 2}'=}0
\end{eqnarray}

\noindent which imply 

\begin{eqnarray}
\bf s'=s_{\rm 1}'=s_{\rm 2}'\qquad{\rm and}\qquad s=s_{\rm 1}=s_{\rm 2}
\nonumber
\end{eqnarray}

\noindent
Expression (\ref{transv}) thus reduces to 

\begin{eqnarray}
&&\int\tilde\varphi_A({\bf s})\tilde\psi^*_A({\bf s}, {\bf s})\tilde\varphi_B({\bf s}')\tilde\psi^*_B({\bf s}',{\bf s}')d^2{\bf s}\,d^2{\bf s}'\cr
&&=\Bigl[\int\tilde\varphi_A({\bf s})\tilde\psi^*_A({\bf s}, {\bf s})d^2{\bf s}\Bigr]\cdot\Bigl[\tilde\varphi_B({\bf s}')\tilde\psi^*_B({\bf s}',{\bf s}')d^2{\bf s}'\Bigr]
\nonumber
\end{eqnarray}

\noindent which shows that the two interactions are localised in the same region in transverse space. 

The overlap of the two amplitudes in Fig.\ref{fig:transverse_interference} restricts considerably the kinematical configurations available to the final state. In fact, the amplitude on the left hand side of the cut in Fig.\ref{fig:transverse_interference} depends, in particular, on $\bf P-Q$. On the other hand, the difference between the two partons' transverse momenta entering the soft blob $\phi_B^*$, in the amplitude on the right hand side of the cut in Fig.\ref{fig:transverse_interference}, is $({\bf P-k-\delta)-(Q-k+\delta)=P-Q}-2\delta$ and the difference between the two partons' transverse momenta entering the soft blob $\phi_A^*$ is $({\bf k+\delta)-(k-\delta)}=2\delta$. So, when $\delta$ is small, the relative momentum of the two lines entering the soft blow $\phi_B^*$ is about ${\bf P-Q}$ and when $2\delta\approx{\bf P-Q}$,  the relative momentum of the two lines entering the soft blow $\phi_A^*$ is about ${\bf P-Q}$: As one may notice, by looking at Fig.\ref{fig:transverse_interference}, the soft blob in the amplitude on the left hand side of the cut limits ${\bf P+Q}$ to values of the order of $1/R$, with $R$ the hadron radius, without putting a similar bound on the difference ${\bf P-Q}$, which is only governed by the hard process. As in the case of the diagonal contribution to the cross section in Fig.\ref{fig:transverse_double}, the difference ${\bf P-Q}$ is, on the contrary, limited to values of the order of $1/R$ by the amplitude on the right hand side of the cut. The final state configurations generated by the two amplitudes in Fig.\ref{fig:transverse_interference} are thus very different and, in a kinematical regime where the single and double parton scattering integrated cross sections are comparable, the effect is a sizable reduction of the contribution of the interference term to the cross section\cite{Calucci:2009ea}. 

One should point out that the same phase space restriction holds for the DPI diagonal term discussed in the previous section\cite{Diehl:2011yj, Markus+Jo}. The reduction of the final state phase-space available to DPIs, as compared to SPIs, originates from the non perturbative scale, of order ${\cal O}(R)$, characterizing the process. By dimensional reasons (c.f.r Eq.\ref{double scattering}) DPIs are thus suppressed at large $p_t^c$ and, conversely, DPIs grow faster, as compared to SPIs, when $p_t^c$ decreases. The c.m. energy and the cut $p_t^c$ determine the lower limit of the initial parton's fractional momenta in the process and thus the initial parton flux. At large c.m. energies and fixed $p_t^c$, the parton flux of DPIs at low $x$ grows about as the square of the parton flux of SPIs, which may compensate the phase space restriction, in such a way that the integrated rate of DPIs can also exceed the integrated rate of SPIs. The flux factor of the interference term grows less rapidly and, in comparison with the diagonal terms, its contribution to the cross section is thus suppressed. 

\section{The simplest Poissonian model}

A simple generalization of the expression of the DPI cross section, where all correlations are neglected, leads to a Poissonian distribution of
MPIs, with  average number depending on the
value of the impact parameter\cite{Ametller:1987ru}, which is a basic feature of most Montecarlo simulation codes of high energy hadronic interactions \cite{Sjostrand:1987su, Butterworth:1996zw, Bahr:2008dy}. 

One may start by 
introducing in the SPI cross section the three dimensional parton density $D(x,b)$, namely
the average number of partons with a given momentum fraction $x$
and with transverse coordinate $b$ (the dependence on flavour and
on the resolution of the process is understood) and making the
simplifying assumption $D(x,b)=G(x)f(b)$, where $G(x)$ is the usual
parton distribution function and $f(b)$ is normalized to one. The SPI
inclusive cross section is thus given by

\begin{eqnarray}
\sigma_S&=&\int_{p_t^c} G_A(x)\hat{\sigma}(x,x')G_B(x') dxdx'\cr
&=&\int_{p_t^c} G_A(x)f_A(b) \hat{\sigma}(x,x')G_B(x')f_B(b-\beta)
         d^2 bd^2\beta dxdx'
\end{eqnarray}

\noindent where integrations are done with the lower limit in the exchanged transverse momenta $p_t^c$. 

The expression allows a simple geometrical
interpretation. Given the large momentum exchange, 
the partonic interaction is localised in the overlap volume of the two
hadrons and one may identify with $b$ and with $b-\beta$ the transverse
coordinates of the two colliding partons, $\beta$ being the
impact parameter of the hadronic collision.

\noindent In the case of a DPI, neglecting all correlations in the multi-parton
distributions, one has 

\begin{eqnarray}
\Gamma(x_1,x_2;r)=G(x_1)G(x_2)F(r)\equiv G(x_1)G(x_2)\int f(b)f(b-r) d^2 b
\end{eqnarray}

\noindent
and, if the two interactions are identical, the DPI cross section $\sigma_D$ in Eq.(\ref{double scattering}) is thus expressed by

\begin{eqnarray}
\sigma_D&=&{1\over 2!}\int_{p_t^c} G_A(x_1)G_A(x_2)f_A(b)f_A(b-r)\hat{\sigma}(x_1,x_1')d^2bdx_1dx_1'\times\nonumber\\
&&\qquad\qquad\times
         \hat{\sigma}(x_2,x_2') G_B(x_1')G_B(x_2')f_B(b')f_B(b'-r)
         d^2b'dx_2dx_2'd^2r\nonumber\\
         &=&{1\over 2!}\int_{p_t^c} G_A(x_1)f_A(b_1)\hat{\sigma}(x_1,x_1')G_B(x_1')f_B(b_1-\beta)d^2b_1dx_1dx_1'\times\nonumber\\
&&\qquad\qquad\times
         G_A(x_2)f_A(b_2)\hat{\sigma}(x_2,x_2') G_B(x_2')f_B(b_2-\beta)
         d^2b_2dx_2dx_2'd^2\beta\nonumber\\
         &=&\int{1\over 2!}\Big(\int_{p_t^c} G_A(x)f_A(b)\hat{\sigma}(x,x')G_B(x')f_B(b-\beta)d^2bdxdx'\Big)^2d^2\beta
\end{eqnarray}

\noindent where the following change of variables has been made: $b_1=b$, $b_1-\beta=b'$, $b_2=b-r$ and $b_2-\beta=b'-r$.

The expression may be readily generalised to the case of the
inclusive cross section of $N$ identical partonic interactions $\sigma_N$:

\begin{eqnarray}
\sigma_N=\int{1\over N!}\Big(\int_{p_t^c} G_A(x)f_A(b)\hat{\sigma}(x,x')G_B(x')f_B(b-\beta)d^2bdxdx'\Big)^Nd^2\beta
\nonumber
\end{eqnarray}

The cross sections $\sigma_N$ may violate unitarity when $p_t^c$ is small. One may however notice that the integrand here above is dimensionless and that it may thus be understood as the probability to have 
$N$-partonic collisions in a inelastic event. The unitarity
problem is thus solved by normalising the integrand. The 
probability $P_N(\beta)$, of having $N$ partonic
interactions in a hadronic collisions with impact parameter $\beta$, is thus given by

\begin{eqnarray}\label{prob}
&&P_N(\beta)\equiv{\bigl(\sigma_SF(\beta)\bigr)^N\over N!} e^{-\sigma_SF(\beta)},\qquad{\rm where}\\
&&\sigma_SF(\beta)\equiv\int_{p_t^c} G_A(x)\hat{\sigma}(x,x')G_B(x')dxdx'\int f_A(b)f_B(b-\beta)d^2b
\nonumber
\end{eqnarray}

By summing all probabilities one obtains in this way the hard cross section
$\sigma_{hard}$, namely the contribution to the inelastic cross
section due to all events with {\it at least} one partonic interaction
with momentum transfer exchange larger than the cutoff $p_t^c$:

\begin{eqnarray}\label{shard}
\sigma_{hard}&=&\sum_{N=1}^{\infty}\int P_N(\beta)d^2\beta =\sum_{N=1}^{\infty}\int d^2\beta{\bigl(\sigma_SF(\beta)\bigr)^N\over N!}
 e^{-\sigma_SF(\beta)}\cr
 &=&\int d^2\beta\Bigl[1-e^{-\sigma_SF(\beta)}\Bigr]
\end{eqnarray}

\noindent
For large $\beta$ the overlap function $F(\beta)$ goes to zero. When $\sigma_S$ is large, $\sigma_{hard}$ thus gives a measure of the size of the overlap region. 

The cross section $\sigma_{hard}$ depends on the choice of the cutoff $p_t^c$. By considering the case of two different cutoffs in transverse momenta, $p_1$ and $p_2$, with $p_1<p_2$ and writing $\sigma_SF(\beta)=T(\beta)$, one identifies in $T$ a softer component $T_S$, when $p_1<p_t<p_2$, and a harder component $T_H$, when $p_2<p_t$, in such a way that $T=T_H+T_S$. One may thus look for the MPI contribution to the total inelastic cross section due to harder component $T_H$. By making in Eq.(\ref{shard}) the replacement

\begin{eqnarray}
T^N\Rightarrow\sum_{K=1}^{N}{{N}\choose{K}}T_H^K\times T_S^{N-K}=(T_H+T_S)^N-T_S^N\nonumber
\end{eqnarray}

\noindent
and, summing over $N$, without considering the color degrees of freedom and even without distinguishing between quarks and gluons, one obtains in this way $\sigma_{hard}(p_2)$, namely the contribution to the inelastic cross
section due to all events with at least one partonic interaction
with momentum transfer exchange larger than $p_2$. Actually: 

\begin{eqnarray}
\sigma_{hard}(p_2)&=&\int d^2\beta\Bigl[\sum_{N=1}^{\infty}\frac{(T_H+T_S)^N}{N!}e^{-(T_H+T_S)}-\sum_{N=1}^{\infty}\frac{T_S^N}{N!}e^{-(T_H+T_S)}\Bigr]\cr
\cr
&=&\int d^2\beta\Bigl[e^{(T_H+T_S)}\times e^{-(T_H+T_S)}-e^{T_S}\times e^{-(T_H+T_S)}\Bigr]\cr
\cr
&=&\int d^2\beta\Bigl[1-e^{-T_H(\beta)}\Bigr]=\int d^2\beta\sum_{N=1}^{\infty}\frac{T_H(\beta)^N}{N!}e^{-T_H(\beta)}
\label{sh}
\end{eqnarray}

\noindent
As defined, $\sigma_{hard}$ thus satisfies a natural consistency requirement: it's insensitive to the softer cutoff $p_1$ and, when replacing $p_t^c$ with $p_2$, the expression in Eq.(\ref{shard}) is the same as the expression in Eq.(\ref{sh}). 

While the hard component of the inelastic cross section is given by a Poissonian distribution of multiple parton interactions, with the average number depending on the impact parameter of the hadronic collision, the inclusive cross sections $\sigma_N$ acquires a well defined meaning in terms of factorial moment of the distribution in multiplicity of partonic collisions. By working out the average number of
collisions one in fact obtains:

\begin{eqnarray}
\langle N\rangle\sigma_{hard}=\int d^2\beta\sum_{N=1}^{\infty}
{N\bigl[\sigma_SF(\beta)\bigr]^N\over N!}
 e^{-\sigma_SF(\beta)}=\int d^2\beta \sigma_SF(\beta)=\sigma_S\nonumber
\end{eqnarray}

\noindent which is precisely the expression of the single scattering inclusive cross section.
More in general, one may write:

\begin{eqnarray}
{\langle N(N-1)\dots(N-K+1)\rangle\over K!}\sigma_{hard}
&=&\int d^2\beta \sum_{N=K}^{\infty}
{N(N-1)\dots(N-K+1)\over K!}P_N(\beta)\nonumber\\
&=&\int d^2\beta {1\over K!}\bigl[\sigma_SF(\beta)\bigr]^K=\sigma_K
\end{eqnarray}

Unitarity corrections thus cancel out in all inclusive cross sections $\sigma_K$, which is an explicit proof of the validity of the cancellation of Abramovsky, Gribov and Kancheli (AGK)\cite{Abramovsky:1973fm} in the actual uncorrelated model of MPIs.

It's worth pointing out a few features:

\noindent - One may define two different sets of MPI cross sections: the inclusive cross sections, given by the factorial moments of the distribution in multiplicity of partonic collisions, and the exclusive cross sections, namely the different contributions to
$\sigma_{hard}$ in Eq.(\ref{shard}). Both sets of cross sections are expressed fully explicitly in terms of the same quantity, the average number of collisions at a fixed impact parameter, $\sigma_SF(\beta)$. 

\noindent - All inclusive cross sections are divergent for $p_t^c\to0$. The relation with the multiplicity of collisions
shows that the divergence is due to the factorial moments. In other words the cause of the divergence is the number of collisions, which become very
large at low $p_t$.

\noindent - While all inclusive cross sections become increasingly
large at low $p_t$, because of the increasingly large number of partonic
interactions, all exclusive cross section, where the number of
hard interactions is kept fixed, become, on the contrary,
smaller and smaller at low $p_t$.

\section{A functional approach}

\par A rather general approach to MPIs is by a functional
formalism\cite{Calucci:1991qq}.

One may introduce the exclusive $n$-parton
distributions $W_n(u_1\dots u_n)$, namely the probabilities to have the hadron in a
configuration with $n$ partons with coordinates $u_i$, which
represent the variables $(b_i,x_i)$, being $b$ the transverse
partonic coordinate and $x$ the corresponding fractional momentum.
The scale for the distributions is given by the cut off
$p_t^c$, that defines the separation between soft and hard
collisions, and the distributions are symmetric in the variables
$u_i$. The generating
functional is defined by:

\begin{equation}{\cal Z}[J]=\sum_n{1\over n!}\int J(u_1)\dots J(u_n)W_n(u_1\dots u_n)
du_1\dots du_n.\end{equation}

\noindent The conservation of the probability implies the
normalization condition ${\cal Z}[1]=1$.

\noindent The probabilities of the various configurations, namely the  exclusive
distributions, are the coefficients of the expansion of ${\cal
Z}[J]$ for $J=0$. The coefficients of the expansion of ${\cal
Z}[J]$ for $J=1$ give the many body densities, i.e. the inclusive
distributions:

\begin{eqnarray}\label{incl.distr}D_1(u)={\delta{\cal Z}\over \delta J(u)}\biggm|_{J=1},\quad
                 \quad\
     D_2(u_1,u_2)={\delta^2{\cal Z}\over \delta J(u_1)\delta J(u_2)}
                  \biggm|_{J=1}\quad\dots\end{eqnarray}

\noindent Correlations, which describe how much the distribution deviates from a
Poissonian, are obtained by the expansion of the logarithm of the generating
functional, ${\cal F}[J]\equiv{\rm ln}{\cal Z}[J]$, for $J=1$:

\begin{eqnarray}{\cal F}[J]&=&\int D_1(u)[J(u)-1]du\cr
&+&\sum_{n=2}^{\infty}{1\over n!}
\int C_n(u_1\dots u_n)\bigl[J(u_1)-1\bigr]\dots
  \dots\bigl[J(u_n)-1\bigr]
du_1\dots du_n
\nonumber
\end{eqnarray}

\noindent  Obviously one has ${\cal F}[1]=0$ and, in the
Poissonian case, $C_n\equiv 0, n\ge 2$.

\noindent  A general
expressions of the semi-hard cross section, which takes into
account of all possible MPIs, is

\begin{eqnarray}\label{sigma hard}\sigma_{hard}=\int d\beta\int&&\sum_n{1\over n!}
  {\delta\over \delta J(u_1)}\dots
  {\delta\over \delta J(u_n)}{\cal Z}_A[J]\nonumber\\
  &\times&\sum_m{1\over m!}
  {\delta\over \delta J'(u_1'-\beta)}\dots
  {\delta\over \delta J'(u_m'-\beta)}{\cal Z}_B[J']\\
&\times&\Bigl\{1-\prod_{i=1}^n\prod_{j=1}^m\bigl[1-\hat{\rm P}_{i,j}(u,u')\bigr]
   \Bigr\}\prod dudu'\Bigm|_{J=J'=0}\nonumber
\end{eqnarray}

\noindent where $\beta$ is the impact parameter between the two
interacting hadrons $A$ and $B$ and $\hat{\rm P}_{i,j}$ is the
probability for the parton $i$ (of hadron $A$) to have a hard
interaction with the parton $j$ (of hadron $B$).

\noindent Analogously to the case of nucleus-nucleus interactions\cite{Bialas:1976ed}, the cross section is
obtained in this way by summing all contributions due to all different
hadronic configurations (the sums over $n$ and $m$) and, for each pair
of values $n$ and $m$, one has a contribution to $\sigma_{hard}$
when at least one hard interaction takes place. The interaction probability is here fully determined by the two body interaction probabilities $\hat{\rm P}_{i,j}$, in such a way that the term in curly brackets in Eq.(\ref{sigma hard}) represents the probability to have at least one interaction.
\noindent In the cross section both disconnected
interactions with $n=m$ and connected interactions, also with $n\neq m$, are included.

\noindent One may focus on multiple disconnected interactions, each initiated by two partons. Only the terms with $n=m$ have thus to be taken into account. One may write the term in curly brackets as:

\begin{eqnarray}S\equiv 1-{\rm exp}\sum_{ij}{\rm ln}(1-\hat{\rm P}_{ij})
=1-{\rm exp}\biggl[-\sum_{ij}\Bigl(\hat{\rm P}_{ij}+{1\over
2}\hat{\rm P}_{ij}\hat{\rm P}_{ij} +\dots\Bigr)\biggr]\end{eqnarray}

\noindent where all repeated indices have to be removed. Only of the first term of the expansion of the logarithm thus contributes and one has to make the following substitution:

\begin{eqnarray}S\Rightarrow 1-{\rm exp}\bigl[-\sum_{ij}\hat{\rm P}_{ij}\bigl]\Rightarrow
  \sum_{ij}\hat{\rm P}_{ij}-{1\over 2}
  \sum_{ij}\sum_{k\not=i,l\not=j}\hat{\rm P}_{ij}\hat{\rm P}_{kl}
  \dots\end{eqnarray}

\noindent The resulting cross section is expressed in a rather compact way:

\begin{eqnarray}\label{compact}\sigma_{hard}(\beta)&=&{\rm exp}(\partial)\cdot {\rm exp}(\partial')
  \Bigl[ 1-{\rm exp}\bigl(-\partial\cdot\hat{\rm P}\cdot\partial'\bigr)\Bigr]
  {\cal Z}_A[J]{\cal Z}_B[J']\Bigm|_{J=J'=0}\cr
  &=&\Bigl[ 1-{\rm exp}\bigl(-\partial\cdot\hat{\rm P}\cdot\partial'\bigr)\Bigr]
  {\cal Z}_A[J]{\cal Z}_B[J']\Bigm|_{J=J'=1}\end{eqnarray}

\noindent where all convolutions are understood. In the simplest non-trivial case all correlations
$C_n$ with $n>2$ can be neglected and the cross section simplifies to:

\begin{eqnarray}\sigma_{hard}(\beta)=
  \Bigl[ 1&-&{\rm exp}\Bigl\{{-\int dudu'{\partial_J}\hat{\rm P}(u,u')
  {\partial_{J'}}}\Bigr\}\Bigr]\cr&\cdot&
  {\rm exp}\biggl\{\int D_A(u)J(u)du+{1\over 2}\int C_A(u,v)J(u)J(v)
         dudv\biggr\}\\
  &\cdot &{\rm exp}\biggl\{\int D_B(u)J'(u)du+{1\over 2}\int C_B(u,v)J'(u)J'(v)
         dudv\biggr\}
\biggm|_{J=J'=0}\nonumber\end{eqnarray}

\noindent which can be worked out explicitly. One 
obtains\cite{Calucci:1991qq}

\begin{eqnarray}\label{exponential}\sigma_{hard}(\beta)=1-{\rm exp}\Bigl[-\sum_n{a_n\over 2}-\sum_n{b_n\over 2n}\Bigr]
\end{eqnarray}

\noindent where

\begin{eqnarray}\label{a_n}
a_n=(-1)^{n+1}\int&& D_A(u_1)\hat{\rm P}(u_1,u_1')C_B(u_1',u_2')
       \hat{\rm P}(u_2',u_2)
       C_A(u_2,u_3)\dots\cr
       &&\qquad\cdots
       \hat{\rm P}(u_n,u_n')D_B(u_n')
  \prod_{i=1}^ndu_idu_i'+A\leftrightarrow B\cr
&&\cr
b_n=(-1)^{n+1}\int &&C_A(u,u_1)\hat{\rm P}(u_1,u_1')C_B(u_1',u_2')
           \dots\\
           &&\qquad\cdots\hat{\rm P}(u_n,u_n')C_B(u_n',u')\hat{\rm P}(u',u)
       dudu'\prod_{i=1}^ndu_idu_i'
       \nonumber
       \end{eqnarray}

Remarkably, {\it also in the general case, where all correlations $C_n$ are taken into account}, by working out the factorial moments of the distribution in the number of collisions, one obtains the same result of the simplest Poissonian model: 

From Eq.(\ref{compact}) one may in fact express the hard
cross section as a sum of MPIs:

\begin{eqnarray}\label{shardc}\sigma_{hard}(\beta)&=&\Bigl[ 1-{\rm exp}\bigl(-\partial\cdot\hat{\rm P}\cdot\partial'\bigr)\Bigr]
  {\cal Z}_A[J]{\cal Z}_B[J']\Bigm|_{J=J'=1}\cr
  &=&\sum_{N=1}^{\infty}{\bigl(\partial\cdot\hat{\rm P}\cdot\partial'\bigr)^N\over{N !} } {\rm e}^{-\partial\cdot\hat{\rm P}\cdot\partial'}
  {\cal Z}_A[J]{\cal Z}_B[J']\Bigm|_{J=J'=1}
\end{eqnarray}

\noindent and the average number of collisions is

\begin{eqnarray}\label{sigmas}\langle N\rangle\sigma_{hard}(\beta)  &=&\sum_{N=1}^{\infty}{N\bigl(\partial\cdot\hat{\rm P}\cdot\partial'\bigr)^{N}\over{N !} } {\rm e}^{-\partial\cdot\hat{\rm P}\cdot\partial'}
  {\cal Z}_A[J]{\cal Z}_B[J']\Bigm|_{J=J'=1}\\
  &=&\partial_{J_1}\cdot\hat{\rm P}\cdot\partial_{J_1'}\sum_{N=0}^{\infty}{\bigl(\partial\cdot\hat{\rm P}\cdot\partial'\bigr)^{N}\over{N !} } {\rm e}^{-\partial\cdot\hat{\rm P}\cdot\partial'}
  {\cal Z}_A[J]{\cal Z}_B[J']\Bigm|_{J=J'=1}\cr
  &=&\bigl(\partial_{J_1}\cdot\hat{\rm P}\cdot\partial_{J_1'} \bigr){\cal Z}_A[J]{\cal Z}_B[J']\Bigm|_{J=J'=1}\cr
  &=&\int D_A(x_1;b_1)\hat{\sigma}(x_1x_1') D_B(x_1';b_1-\beta)dx_1dx_1'd^2b_1\cr
  &\equiv&\sigma_S(\beta),\nonumber
  \end{eqnarray}

  \noindent where $\hat{\sigma}(x_1x_1')$ is the parton-parton cross section, integrated with $p_t>p_t^c$. Given the localization of the interactions in transverse space,  the parton-parton interaction probability has in fact been represented as a $\delta$-function of the transverse coordinates: $\hat{\rm P}(u,u')=\hat{\sigma}(x, x')\delta({\bf b}-{\bf b}')$. Analogously 

\begin{eqnarray}\label{sigmad}{\langle N(N-1)\rangle\over2!}\sigma_{hard}(\beta)&=&{1\over2!}\int D_A(x_1x_2;b_1 b_2)\hat{\sigma}(x_1x_1')\hat{\sigma}(x_2x_2')\\&&\quad\quad\times   D_B(x_1' x_2';b_1-\beta, b_2-\beta)dx_1dx_1'd^2b_1 dx_2dx_2'd^2b_2\cr
&\equiv&\sigma_D(\beta)\nonumber
\end{eqnarray}

\noindent and, in general,

\begin{eqnarray}\label{sigmak}&&\!\!\!\!\!\!\!\!\!\!\!{\langle N(N-1)\dots(N-K+1)\rangle\over K!}\sigma_{hard}(\beta)\cr
  &&\qquad\qquad={1\over K!}\int D_A(x_1 \dots x_K;b_1 \dots b_K)\hat{\sigma}(x_1x_1')\dots\hat{\sigma}(x_Kx_K')\\
  &&\qquad\quad\qquad\quad\times  D_B(x_1' \dots x_K';b_1-\beta \dots b_K-\beta)dx_1dx_1'd^2b_1\dots dx_Kdx_K'd^2b_K \nonumber\\
&&\qquad\qquad\equiv\sigma_K(\beta)\nonumber
\end{eqnarray}

\noindent which shows that {\it when considering the case, where all connected interactions are neglected and where each partonic interaction is initiated by two partons}, for any choice of multiparton distributions, the
inclusive cross sections are given by the factorial moments of the
distribution in the number of partonic collisions. 

The validity of the cancellation of AGK\cite{Abramovsky:1973fm} is thus proved explicitly in the actual model of MPIs on rather general grounds.

\section{Exclusive cross sections, sum rules}

With the simplifying assumptions here above, in $pp$ collisions the inclusive cross sections are thus given by the factorial moments of the
distribution in the number of MPIs. The most basic information on the distribution in the number of collisions, the average number, corresponds to the
single scattering inclusive cross section of the pQCD parton model and, analogously, the $K$-parton scattering inclusive cross
section $\sigma_K$ corresponds to the $K$th factorial moment of the distribution in the number
of collisions. 

As already pointed out, a way alternative to the set of factorial moments, to provide the whole
information of the distribution, is represented by the set of the
different terms of the probability distribution of multiple
collisions. In addition to the set of the
inclusive cross sections $\sigma_K$, one may thus consider the set of the
exclusive cross sections $\tilde\sigma_N$, which correspond to the different terms of the probability distribution and which represent the cross sections where one selects the events where $\it only$ a given number $N$ of collisions are present\cite{Calucci:2009sv, Seymour:2013sya}. The following relations thus hold:

\begin{eqnarray}\label{sum rules}
\sigma_{hard}\equiv\sum_{N=1}^{\infty}\tilde\sigma_N,\qquad\sigma_K\equiv\sum_{N=K}^{\infty}\binom{N}{K}\tilde\sigma_N
\end{eqnarray}

\noindent which may be also understood as a set of sum rules connecting the inclusive and the exclusive cross sections.

While the non perturbative input to the inclusive cross section $\sigma_K$ is given by the $K$-parton distributions of the hadron structure, as implicit in Eq.s (\ref{shard}) and (\ref{shardc}), the non-perturbative input to the exclusive cross sections is given by an infinite set of multi-parton distributions. On the other hand, the request of being in a perturbative regime limits the number of partonic collisions and the sum rules in Eq.(\ref{sum rules}) can be saturated by a few terms, in such a way that the exclusive cross sections can be expressed by finite combinations of inclusive cross sections. 

A particular case, already considered in the previous section, where all correlations
$C_n$ with $n>2$ are negligible, can be worked out fully explicitly. The exponential in Eq.(\ref{exponential}) represents the probability of no interaction at a given impact parameter $\beta$ and all exclusive cross sections can be obtained from the argument of the exponential. 

Consider the interaction probability 

\begin{eqnarray}1-\prod_{i,j=1}^N(1-\hat{\rm P}_{ij})
\end{eqnarray}

\noindent where $N$ is the maximal number of possible partonic interactions between two given initial partonic configurations and each index assumes a given value
only once, in such a way that possible connected interactions are not
included. The probability of having only a single interaction is 

\begin{eqnarray}\Biggl(-{\partial\over\partial g}\Biggr)\prod_{i,j=1}^N(1-g\hat{\rm P}_{ij})\Bigg|_{g=1}=\sum_{kl}\hat{\rm P}_{kl}\prod_{ij\neq kl}^N(1-g\hat{\rm P}_{ij})\Bigg|_{g=1},\end{eqnarray}

\noindent the probabilities of a double and of a triple
interaction are

\begin{eqnarray}{1\over 2!}\Biggl(-{\partial\over\partial g}\Biggr)^2\prod_{i,j=1}^N(1-g\hat{\rm P}_{ij})\Bigg|_{g=1}&=&{1\over 2!}\sum_{kl}\sum_{rs}\hat{\rm P}_{kl}\hat{\rm P}_{rs}\prod_{ij\neq
(kl,rs)}^N(1-g\hat{\rm P}_{ij})\Bigg|_{g=1}\\
{1\over 3!}\Biggl(-{\partial\over\partial
g}\Biggr)^3\prod_{i,j=1}^N(1-g\hat{\rm P}_{ij})\Bigg|_{g=1}&=&{1\over
3!}\sum_{kl}\sum_{rs}\sum_{tu}\hat{\rm P}_{kl}\hat{\rm P}_{rs}\hat{\rm P}_{tu}\prod_{ij\neq
(kl,rs,tu)}^N(1-g\hat{\rm P}_{ij})\Bigg|_{g=1},\nonumber\end{eqnarray}

\noindent while the corresponding expressions for the exclusive
cross sections are

\begin{eqnarray}\label{exclusiveX}\Biggl(-{\partial\over\partial
g}\Biggr)\esp{-X(g)}\Bigg|_{g=1}&=&X'(g)\esp{-X(g)}\Bigg|_{g=1}\nonumber\\
{1\over2!}\Biggl(-{\partial\over\partial
g}\Biggr)^2\esp{-X(g)}\Bigg|_{g=1}&=&{1\over2!}\Bigl\{[X'(g)]^2-X''(g)\Bigr\}\esp{-X(g)}\Bigg|_{g=1}\\
{1\over3!}\Biggl(-{\partial\over\partial
g}\Biggr)^3\esp{-X(g)}\Bigg|_{g=1}&=&{1\over3!}\Bigl\{X'''(g)+[X'(g)]^3-3X'(g)X''(g)\Bigr\}\esp{-X(g)}\Bigg|_{g=1},\nonumber
\end{eqnarray}

\noindent where $X={1\over2}(\sum a_n+\sum b_n/n)$ depends on $g$ through $a_n$ and $b_n$  (cfr Eq.\ref{a_n}) by the replacement $\hat{\rm P}\to g\hat{\rm P}$.
If however one is interested in expression where some kinematical variables are free, i.e. not integrated, the substitution to be performed is: 
$\hat{\rm P}(u,u')\to g(u,u')\hat{\rm P}(u,u')$ and the subsequent derivatives becomes functional derivatives $\delta/\delta g(u,u')$; we keep symbols like $X',\;X''$ also for the functional derivatives.

It's convenient to expand $X$ and its derivatives in the
number of elementary collisions

\begin{eqnarray}
X&=&X_1+X_2+X_3+\dots\,\,\,\,\,\,\,{\rm , where:}\nonumber\\
&&\nonumber\\
X_1&=&\int D_A(u)\hat{\rm P}(u,u') D_B(u')dudu'\nonumber\\
X_2&=&-{1\over2}\Bigl[\int D_A(u_1)\hat{\rm P}(u_1,u_1') C_B(u_1',u_2')
       \hat{\rm P}(u_2',u_2)D_A(u_2)\prod_{i=1}^2du_idu_i'+A\leftrightarrow B\Bigr]\nonumber\\
&&\qquad-{1\over2}\int C_A(u_1,u_2)\hat{\rm P}(u_1,u_1') C_B(u_1',u_2')
       \hat{\rm P}(u_2',u_2)\prod_{i=1}^2du_idu_i'\\
X_3&=&\int D_A(u_1)\hat{\rm P}(u_1,u_1') C_B(u_1',u_2')
       \hat{\rm P}(u_2',u_2)C_A(u_2,u_3)
       \hat{\rm P}(u_3,u_3')\nonumber\\
&&\qquad\times D_B(u_3')\prod_{i=1}^3du_idu_i'\nonumber
\end{eqnarray}

It should be pointed out that, when expanding at a fixed order in powers of $\hat{\rm P}$, one could easily introduce the triple correlations $T$ in the multi-parton distributions by including in $X_3$ terms like $\int D_A(u_1)\hat{\rm P}(u_1,u_1') D_A(u_2)\hat{\rm P}(u_2,u_2') D_A(u_3)\hat{\rm P}(u_3,u_3')T_B(u_1',u_2',u_3')\prod du_idu_i'$ and similar combinations involving the double correlation $C$\cite{Calucci:2009sv}. 

The derivatives at $g=1$ can be worked out\cite{Calucci:2009sv}. By expanding Eq.(\ref{exclusiveX}) and its derivatives in the number of
elementary collisions and disregarding all terms beyond the third order in $\hat P$, one obtains the following expressions:

\begin{eqnarray}
\tilde\sigma_1'&=&(X'_1+X'_2+X'_3)(1-X_1-X_2+X_1\cdot X_1/2)\nonumber\\
2\times\tilde\sigma_2''&=&(X'_1\cdot X'_1+2X'_1\cdot X'_2-X_2''-X_3'')(1-X_1)\\
3\times\tilde\sigma_3'''&=&{1\over2}(X'''_3+X_1'\cdot X_1'\cdot X_1'-3X_1'\cdot X_2'')\nonumber
\end{eqnarray}

\noindent where $\tilde\sigma_1'$ etc. are the exclusive cross sections, differential in the different partonic variables.  

The integrated exclusive cross sections are

\begin{eqnarray}
\tilde\sigma_1&=&X_1-X_1^2-X_1X_2+X_1^3/2+2X_2-2X_2X_1+3X_3\nonumber\\
2\times\tilde\sigma_2&=&X_1^2+4X_1X_2-2X_2-6X_3-X_1^3+2X_1X_2\\
3\times\tilde\sigma_3&=&3X_3+(X_1)^3/2-3X_1X_2\nonumber
\end{eqnarray}

\begin{figure}[htp]
\centering
\includegraphics[width=13cm]{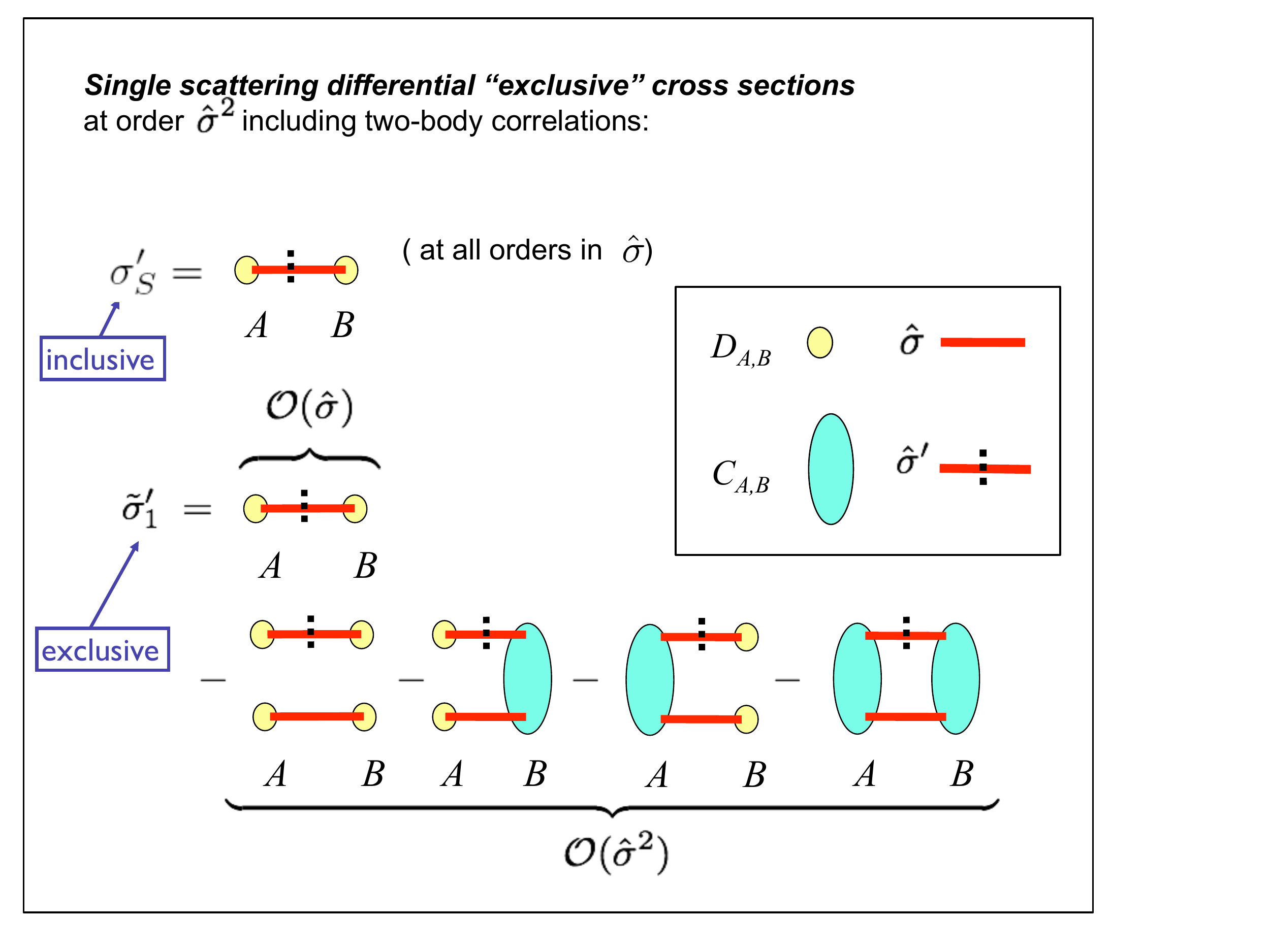}
\vspace{0cm}
\caption{Symbolic representation of the inclusive and of the exclusive cross sections $\sigma_S'$ and $\tilde\sigma_1'$ at order $\hat\sigma^2$ (see the main text).}
\label{fig:correlations}
\end{figure}

\vspace{-.1in}
A representation of the inclusive cross section $\sigma_S'$, at all orders in the number of elementary collisions  $\hat\sigma$, and of the exclusive cross sections $\tilde\sigma_1'$, at order $\hat\sigma^2$, is shown in Fig.\ref{fig:correlations}. The yellow circles represent the inclusive one-body parton distribution, the green oval the two body parton correlation, the red line the elementary interaction, differential when the red line is crossed by black points.

\noindent The sum rules of Eq.(\ref{sum rules}) are satisfied as follows:

\begin{eqnarray}\label{exclusive}
\tilde\sigma_1+2\times\tilde\sigma_2+3\times\tilde\sigma_3=&&X_1\nonumber\\&&-X_1^2+2X_2+X_1^2-2X_2-X_1X_2+X_1^3/2-2X_2X_1+3X_3\nonumber\\
&&+4X_1X_2-6X_3-X_1^3+2X_1X_2+3X_3+(X_1)^3/2-3X_1X_2\nonumber\\
=&&X_1\equiv\sigma_S
\\
2\times\tilde\sigma_2+6\times\tilde\sigma_3
=&&X_1^2-2X_2\nonumber\\
&&+4X_1X_2-6X_3-X_1^3+2X_1X_2+6X_3+(X_1)^3-6X_1X_2\nonumber\\
=&&X_1^2-2X_2\equiv2\times\sigma_D
\nonumber\\
6\times\tilde\sigma_3=&&6X_3+(X_1)^3-6X_1X_2
\equiv3!\times\sigma_T\nonumber
\end{eqnarray}

\noindent where $\sigma_S$, $\sigma_D$ and $\sigma_T$ are respectively the single, double and triple parton scattering inclusive cross sections. Explicitly

\begin{eqnarray}
\sigma_S&=&X_1=\int D_A\hat{\rm P}D_B\nonumber\\
\sigma_D&=&{1\over2}[X_1^2-2X_2]={1\over2}\Bigl[\int D_A\hat{\rm P}D_B\cdot D_A\hat{\rm P}D_B+\int D_A\hat{\rm P}C_B\hat{\rm P}D_A\nonumber\\
&&\qquad\qquad\qquad\qquad\qquad+\int D_B\hat{\rm P}C_A\hat{\rm P}D_B+\int C_A\hat{\rm P}C_B\hat{\rm P}\Bigr]\nonumber\\
&&\qquad\qquad\quad={1\over2}\int[D_AD_A+C_A]\hat{\rm P}\hat{\rm P}[D_BD_B+C_B]
\end{eqnarray}

\noindent where $[DD+C]\equiv D_2$ is the two body parton
distribution, as defined in Eq.(\ref{incl.distr}) and the arguments of the two $\hat{\rm P}$s here above are, of course, different. An analogous expression may be written for $\sigma_T$.

The relations (\ref{exclusive}) may be inverted

\begin{eqnarray}\label{tildesigmas}
\tilde\sigma_1&=&\sigma_S-2\sigma_D+3\sigma_T\nonumber\\
\tilde\sigma_2&=&\sigma_D-3\sigma_T\\
\tilde\sigma_3&=&\sigma_T\nonumber
\end{eqnarray}

\noindent which allow expressing the scale parameters characterising the double and triple parton collisions in terms of the single scattering inclusive cross section $\sigma_S$ and of the exclusive cross sections $\tilde\sigma_1$ and $\tilde\sigma_2$:

\begin{eqnarray}\label{sdt}
\sigma_D&=&\sigma_S-\tilde\sigma_1-\tilde\sigma_2={1\over2}{\sigma_S^2\over\sigma_{\rm eff}}\nonumber\\
\sigma_T&=&{1\over3}(\sigma_S-\tilde\sigma_1-2\tilde\sigma_2)={1\over6}\sigma_S^3{1\over\tau\sigma_{\rm eff}^2}
\end{eqnarray}

\noindent where the scale factor of the triple parton scattering cross section has been characterised by the dimensionless parameter $\tau$\cite{dEnterria:2016ids, Snigirev}.

\section{Concluding remarks}

The present model of MPIs is characterised by different non trivial features and its extension to a more general case would represent an important step towards a comprehensive understanding of MPI dynamics. In the model one can in fact prove that:  

\noindent
- The cancellation of AGK\cite{Abramovsky:1973fm} holds for each MPI inclusive cross section, both in the uncorrelated case and when correlations in the many-parton distributions are taken into account, Eq.s (\ref{sigmas}, \ref{sigmad}, \ref{sigmak}). In both cases the MPI inclusive cross sections are thus given by the factorial moments of the distribution in multiplicity of the partonic interactions. 

\noindent
- A consequence is that each MPI inclusive cross section can be safely evaluated at a given order in the number of partonic collisions, since unitarity corrections are not going to spoil the calculation.

\noindent
- In addition to the inclusive cross sections one may introduce the exclusive cross sections, corresponding to the case where only a given number of hard interactions take place. Inclusive and exclusive cross sections are linked by sum rules, Eq.(\ref{sum rules}), which allow to evaluate also the exclusive cross sections, at a given order in the number of partonic collisions, $e.g.$ Eq.(\ref{tildesigmas}).

\noindent
- Inclusive and exclusive cross sections are measured independently and their comparison can thus provide an additional handle for the determination of the non-perturbative parameters, which characterise the MPIs, $e.g.$ Eq.(\ref{sdt}), and the related unknown non-perturbative properties of the hadron structure. 

We like to end showing by an example what we may learn on correlations from high-energy proton-deuteron collisions. 

As already noted, a simple and efficient tool to study many-parton dynamics is given by the effective cross section, Eq.(\ref{double scattering}), $\sigma_{\rm eff}={\sigma_S}^2/(2\cdot \sigma_D)$ and its generalisations. 
 Although $\sigma_{\rm eff}$ is related to the size of the hadron, there are possible concurrent effects, which can be exemplified by two extreme situations: 

\par\noindent
1- Configurations with high multiplicity are frequent so many double (and multiple) collisions are produced, $e.g.$ the parton number follows a negative-binomial distribution instead of a Poissonian distribution, then
 $\sigma_{\rm eff}$ becomes small.
 
 \par\noindent
2- Partons are strictly correlated so that if a pair collides, another pair also collides. Again $\sigma_{\rm eff}$ becomes small.
\par
Observing multiple collisions both on free nucleons and on nucleons bound in light nuclei helps in separating the different form of correlations:
Consider a double hard scattering in proton-deuteron collision\cite{Calucci:2010wg}. There are events where the projectile interacts twice with one nucleon of the target, the other bound nucleon being a spectator, so nothing new may be learned, and there are events where the projectile interacts with both bound nucleons. In the latter case, together with the spatial correlations of the partons, the deuteron wave function plays a relevant role and, since the deuteron wave function is well known, a new window may be opened on the actual content of parton pairs in the nucleon.

\end{document}